\documentclass[dvipdfmx]{ptephy}

\preprintnumber{arXiv:1605.07339} 
\usepackage{bm}

\begin{document}

\title{
Quark confinement potential examined by
excitation energy of the $\Lambda_{c}$ and $\Lambda_{b}$ baryons in 
a quark-diquark model
}


\author{\name{Daisuke Jido}{\ast}, and
\name{Minori Sakashita}{}}
\address{
\affil{}{Department of Physics,
Tokyo Metropolitan University,  1-1 Minami-Osakwa, 
Hachioji, Tokyo, 192-0397, Japan, \email{jido@tmu.ac.jp}}}



\begin{abstract}%

The possibility to have diquark configuration in heavy baryons, such as $\Lambda_{c}$ and $\Lambda_{b}$, 
is examined by a nonrelativistic potential model with a heavy quark and a light scalar diquark. 
Assuming that the $\Lambda_{c}$ and $\Lambda_{b}$ baryons are composed of the heavy quark 
and the point-like scalar-isoscalar $ud$ diquark, we solve the two-body Schr\"odinger equation with 
the Coulomb plus linear potential and obtain the energy spectra for the heavy baryons. 
Contrary to our expectation, it is found that the potential determined by the quarkonium spectra fails to 
reproduce the excitation spectra of the $\Lambda_{c}$ and $\Lambda_{b}$ in the quark-diquark 
picture, while the $\Lambda_{c}$ and $\Lambda_{b}$ spectra is reproduced with a half strength 
of the confinement string tension than for the quarkonium. The finite size effect of the diquark 
is also examined and it is found that introduction of a finite size diquark would resolve the failure 
of the spectrum reproduction. 
The $\Xi_{c}$ excitation energy is also calculated and is found to be smaller than $\Lambda_{c}$
in the quark-diquark model. This is not consistent with the experimental observation. 
 
\end{abstract}


\maketitle

\section{Introduction}
Hunts for fundamental correlations in strongly interacting quarks and gluons
inside hadron can be a hint to understand the hadron structure and 
the confinement mechanism. The constituent quarks are considered to be
good effective degrees of freedom for the hadron structure, 
especially explaining the nucleon magnetic moment based on the spin flavor 
symmetry of the light quarks and low lying energy spectra of the heavy quarkonia
by quark excitations in the linear-plus-Coulomb confinement potential.   
The diquark as two quark
pair correlation in hadron can be also a strong candidate of hadron 
constituent~\cite{Ida:1966ev,Lichtenberg:1967zz}.
So far, the existence of diquark correlation inside hadrons has been suggested
by phenomenological findings in baryon spectroscopy, weak non-leptnic decays,
parton distribution functions and fragmentation functions,
and especially the scalar diquark with flavor, spin and color antisymmetric configurations, 
which is so-called good diquark,
is expected to have the most attractive correlation~\cite{Jaffe:2004ph}.
The strong correlation in the scalar diquark is obtained by
one-gluon exchange and by instanton induced interactions, and 
has been found in Lattice calculations~\cite{Hess98,Babich,Orginos,Alexandrou}.
The strong correlation can be a remnant of
the Pauli-G\"ursey symmetry in two-color QCD, in which quark-quark
interactions for massless quarks are exactly same in strength  
as quark-antiquark interactions. In three-color QCD, this symmetry 
is explicitly broken and the quark-quark correlations get
less attractive than the quark-antiquark correlation. 

To investigate the nature of the confinement potential, we focus on the 
excitation spectra of the heavy baryons. 
So far, many works have been done for baryons spectra in quark-diquark models. 
In Ref.~\cite{Goldstein:1979wba}, the mass spectra of $p$ and $d$ excited states of 
non-strange baryons were investigated in a diquark-quark model based on the SU(6)$\otimes$O(3)
classification~\cite{Lichtenberg:1981pp} in special attentions of the fine structure by spin-orbit interactions.
It was found that the detailed confinement mechanism is irrelevant for the fine
structure of the intramultiplet splitting. In Ref.~\cite{Lichtenberg:1982jp}, the ground 
states of spin 3/2 baryons were investigated in a relativistic formulation.
In Ref.~\cite{Liu:1983us}, radial and orbital excitations of baryons were calculated
in a diquark-quark model with a confinement potential reproducing meson spectra,
and detailed analyses were given for light flavor baryon spectra. 
Recently, in Ref.~\cite{Kim:2011ut}, the ground state masses of $\Lambda$, $\Lambda_{c}$
and $\Lambda_{b}$ were calculated in a diquark QCD sum rule, in which 
a scalar diquark is explicitly considered as a fundamental field in operator product expansion,
and the calculation successfully reproduced the observed $\Lambda$'s masses with 
a ``constituent" diquark mass 0.4 GeV, satisfying the standard criteria for 
the QCD sum rule to work well. 

The system of a heavy quark and a diquark has an advantage to investigate the light diquark properties. 
As reported in Ref.~\cite{Jaffe76-1}, in meson wavefunctions, quark-antiquark component 
are dominated and diquark configuration are rather suppressed. This is because  
the diquark correlation is weaker than the quark-antiquark correlation and,
once there exists a antiquark closed to a diquark, the diquark could easily
fall apart and form a quark-antiquark pair.
In light baryons, diquark configuration could play an important role for the baryon structure, 
but rearrangement of the diquark would be also important due to 
the symmetry among the light quarks. In systems of a heavy quark and two light quarks
forming a heavy baryon, thanks to asymmetry between the light quark and the heavy quark, 
a diquark of two light quarks could easily emerge in the baryon.

In this brief note, taking the quark-diquark picture with a point-like diquark,
we report on the excitation energy of the $\Lambda_{c}$ and $\Lambda_{b}$ baryons
using the linear-plus-Coulomb confinement potential 
suitable for the quarkonium spectra.  Here we will find a puzzle that
the string tension in the confinement potential for quark and diquark systems
should be half as strong as that for quark and antiquark systems
to reproduce the $\Lambda_{c}$ and $\Lambda_{b}$ excitation energies.
This is against the universality of the confinement force. 
To understand the overestimation of these excitation energies,
we will examine the finite size effect of the diquark, and find 
that introduction of the diquark size could be one of the solutions 
of the above puzzle.  We will also find that 
the excitation energy of the $\Xi_{c}$ baryon should be smaller that 
that of $\Lambda_{c}$ in the quark-diquark picture, but experiments
tell us that it is opposite. These puzzles should be solved when one takes
the quark-diquark picture for the heavy baryons.

\section{Formulation}
We describe the heavy baryons composed of one heavy and two light quarks
by the diquark model in which the light quarks form the scalar diquark with antisymmetrized  
flavor, color 
anti-triple $\bar{\bf 3}$ and spin-parity $0^{+}$. We regard the heavy quark (a charm or 
bottom quark) and the scalar diquark as elementary particles and take a quark potential 
model in the non-relativistic formulation. We also assume that the diquark is a point-like particle.
In the center of mass frame,  
the Hamiltonian of the system including the rest masses of the quark and diquark 
is written as
\begin{equation}
   H = m_{h} + m_{d} + \frac{p^{2}}{2\mu} + V(r)
\end{equation}
with the heavy quark mass $m_{h}$, the diquark mass $m_{d}$, 
the momentum of the relative motion $p$ 
and the reduced mass $\mu$.  
We use the $c$ and $b$ quark masses as $m_{c} = 1.5$ GeV/$c^{2}$ and $m_{b}=4.0$/$c^{2}$.
The interaction between the heavy quark and 
the scalar diquark is assumed to be spherical and written as a Coulomb plus linear form~\cite{Eichten:1974af}
\begin{equation}
   V(r) = -\frac{4}{3} \frac{\alpha}{r} \hbar c + k r + V_{0}  \label{pot}
\end{equation}
with three constant parameters $\alpha$, $k$ and $V_{0}$. The Coulomb part proportional to $r^{-1}$
expresses the asymptotic nature of the strong interaction in short distance,
and the linear part, $kr$, represents the confinement potential with the string tension $k$ of the flux tube. 
The constant $V_{0}$ adjusts the absolute 
value of the potential energy, turning out to be an irrelevant parameter in the present analysis 
because we are interested in the excitation energies of quark-diquark systems.
It is known that the potential $V(r)$ given in Eq.~(\ref{pot}) reproduces the charmonium 
and bottonium spectra well with appropriate spin-spin and spin-orbit interactions~\cite{Mukherjee:1993hb}.
The scalar diquark of our interest has an anti-triplet color charge and 
it is same as the color charge of antiquark. 
Because the color electric force should be independent of the quark flavor in the first approximation 
and be determined by the color charges, we may make good use of the potential determined 
by the quarkonium spectra for the quark-diquark system. 

The energy spectrum and the wave function of the quark-diquark system can be 
obtained by solving the Sch\"odinger equation for the radial variable of the relative motion:
\begin{equation}
   - \frac{\hbar^{2}}{2 \mu} \frac{d^{2} \chi_{\ell}(r)}{dr^{2}} 
   + \left[ V(r) + \frac{\ell(\ell+1)\hbar^{2}}{2\mu r^{2}}\right] \chi_{\ell}(r)
   = E_{\ell} \chi_{\ell}(r),  \label{sch}
\end{equation}
where $\mu$ is the reduced mass of the heavy quark and scalar diquark
and $\chi_{\ell}(r) \equiv R_{\ell}(r) / r$ is the radial wave function 
for the orbital angular momentum $\ell \hbar$. 
The mass of the bound system for the heavy quark and the scalar diquark is given by $M = m_{h} + m_{d} + E_{\ell}$.
Because this Sch\"odinger equation is a one-body 
equation after separating out the center of mass motion, it is simple enough to be solved numerically. 
We use the state label defined by $n_{r} \ell$ with the quantum numbers,  $n_{r}$ and $\ell$, for the 
radial excitation and the orbital angular momentum, respectively. We are of great interest in the excitation 
pattern of the quark-diquark system in the confinement potential. We focus on the energy spectrum 
without considering the fine structure caused by the spin-orbit interaction. 
The fine structure is not sensitive to the details of the confinement mechanism~\cite{Goldstein:1979wba}.
We may refine our model quantitatively, for instance, by fine-tuning quark and diquark masses,
or introducing scale dependence of the Coulomb part of the confinement potential.
These are important next steps to be done. Here we are of great interest in the global 
structure of the excitation spectra of the heavy baryon. Thus, as a first step,
we ignore these corrections and spin-orbit interaction. 

\section{Results}

In this section, we present our results of the energy spectra obtained by solving 
the Schr\"odinger equation~(\ref{sch}) with the confinement potential~(\ref{pot}).
First we show the charmonium and bottonium spectra to confirm that the confinement potential
with $\alpha=0.4$ and $k=0.9$ GeV/fm reproduces the observed major excitation well. 
We will also see that with the same parameters the model works for the heavy mesons, 
$D$, $D_{s}$, $B$, $B_{s}$, with the light quark mass $m_{q}=0.3$~GeV/$c^{2}$ and
the strange quark mass $m_{s}=0.5$ GeV/$c^{2}$. Next, finding that the confinement potential 
suitable for the meson spectra does {\it not} work in the quark-diquark systems
to reproduce the excitation energies of $\Lambda_{c}$ and $\Lambda_{b}$,
we search for the appropriate parameter set to reproduce the $\Lambda_{c}$ and $\Lambda_{b}$
spectra. We will calculate the $\Xi_{c}$ and $\Xi_{b}$ excitation energies by
introducing the strange scalar diquark composed of a strange quark and a light quark. 
We also estimate the size of the decay width of the excited state of $\Lambda_{c}$ 
based on the quark-diquark picture
by applying Gamov theory for the nuclear $\alpha$ decay. 
Finally, we examine the finite size effect of the diquark. 

\subsection{Quarkonium spectra}

\begin{figure}
 \begin{center}
  \includegraphics[width=9cm]{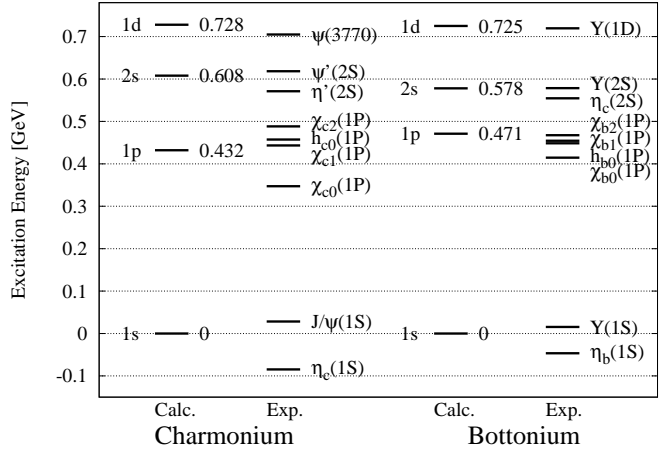}
 \end{center}
 \caption{Excitation spectra of charmonium and bottonium in our model 
 with $\alpha = 0.4$, $k = 0.9$ GeV/fm, $m_c=1.5$ GeV/$c^{2}$ and $m_{b}=4.0$ GeV/$c^{2}$. 
 In the left line for each quarkonium the calculated values are shown with the state label and a number. 
 The numbers are excitation energies in units of GeV measured from the ground state energy.  
 The states are labeled by $n_{r} \ell$ with the radial quantum 
 number $n_{r}$ and the angular momentum quantum number $\ell$. 
 The calculation is done without spin-spin nor spin-orbit interaction.
 In the right line for each quarkonium we show the excitation energies of the observed quarkonia
 measured from the spin weighted average of the $1s$ ground state energies.
 The experimental data are taken from Particle Data Group~\cite{PDG}.
 }
 \label{fig:quarkonia}
\end{figure}

Let us first see that the confinement potential~(\ref{pot}) with $\alpha=0.4$ and $k=0.9$ GeV/fm
works well for the charmonium and bottonium spectra~\cite{Quigg:1979vr}. 
In Fig.~\ref{fig:quarkonia}, we plot the excitation
energy spectra of the charmonium and bottonium measured from each ground state. 
The calculation is done with the potential given in Eq.~(\ref{pot}) with $\alpha= 0.4$ and $k=0.9$ GeV/fm
without spin-spin nor spin-orbit interaction. The quark masses are fixed as $m_{c} = 1.5$ GeV/$c^{2}$ for 
the charm quark and $m_{b} = 4.0$ GeV/$c^{2}$ for the bottom quark. 
To compare the calculation without the spin-spin and spin-orbit interaction, 
the excitation energies of the observed charmonia is measured from 
the spin wighted average of the $\eta_{c}$ and $J/\psi$ masses, $(m_{\eta_{c}}+3m_{J/\psi})/4$, 
which removes the effect of the spin-spin interaction perturbatively,
since the matrix element of the spin-spin operator $\vec s_{1} \cdot \vec s_{2}$
for the quark-antiquark system with total spin $S$ and the angular momentum $\ell=0$
is calculated as
\begin{equation}
  2 \langle S | \vec s_{1} \cdot \vec s_{2} |S \rangle =  S(S+1) - \frac{3}{2}
  = \left\{ \begin{array}{rl} -\frac{3}{2} \quad & {\rm for\ } S=0,  \\  + \frac{1}{2} \quad & {\rm for \ } S=1,
    \end{array} \right. 
\end{equation}
In the same way, the excitation energies of the observed bottonia are measured from 
the spin weighted average of $\eta_{b}$ and $\Upsilon$, $(m_{\eta_{b}}+3m_{\Upsilon})/4$.
Figure~\ref{fig:quarkonia} shows that the potential (\ref{pot}) with $\alpha = 0.4$ and $k = 0.9$ GeV/fm 
successfully reproduces the size of the major excitations (radial and orbital excitations) both 
for charmonium and bottonium. 

\begin{figure}
 \begin{center}
  \includegraphics[width=9cm]{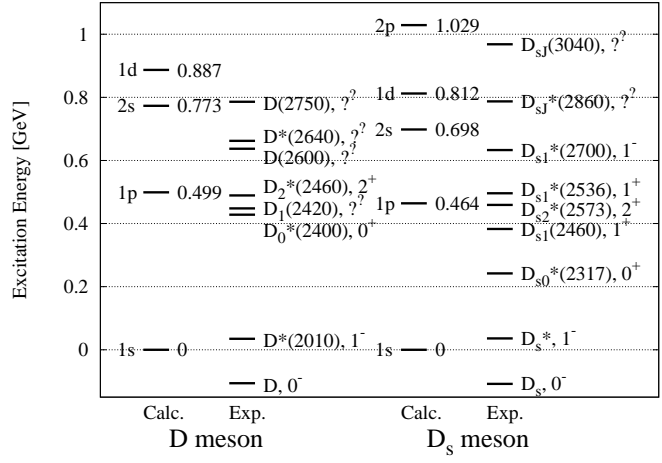}
 \end{center}
 \caption{
 Same as Fig.~\ref{fig:quarkonia} for the $D$ and $D_{s}$ mesons with $m_q=0.3$ GeV/$c^{2}$ 
 and $m_{s}=0.5$ GeV/$c^{2}$. 
 For the $D$ meson, we show the spectrum for the charged $D$ meson. 
 The experimental data are taken from Particle Data Group~\cite{PDG}. 
}
 \label{fig:Dmeson}
\end{figure}

\begin{figure}
 \begin{center}
  \includegraphics[width=9cm]{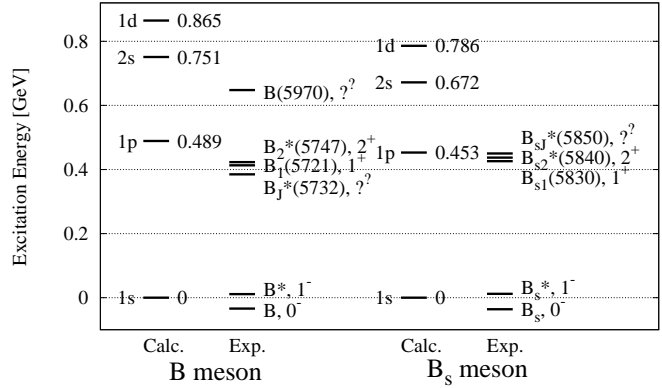}
 \end{center}
 \caption{
 Same as Fig.~\ref{fig:Dmeson} for the $B$ and $B_{s}$ mesons.
The experimental data are taken from Particle Data Group~\cite{PDG}. 
}
 \label{fig:Bmeson}
\end{figure}

We also calculate the excitation energies of the $D$ and $D_{s}$ mesons, which are composed of a charm quark and
a light (up, down or strange) quark, by using the same potential parameters and the light quark mass 
$m_{q} = 0.3$ GeV/$c^{2}$ and the strange quark mass $m_{s}=0.5$ GeV/$c^{2}$. 
As shown in Fig.~\ref{fig:Dmeson}, 
we obtain 0.50 GeV and 0.77 GeV for the $D$ meson excitation energies of the $1p$ and $2s$ states of  
from the ground state ($1s$), respectively. These values should be compared with the experimental data.
The experimental excitation energies are shown again from the spin-wighted averaged value
of the ground states with $\ell=0$. 
Because the excited states of the $D$ and $D_{s}$ mesons are unstable against the strong interaction with
large decay widths and the quantum numbers of some states are not well-known yet,
the detailed comparison is somewhat difficult. Nevertheless, it is observed that 
the excitation energies of the first excited states of the $D$ meson are around 0.45 to 0.57 GeV.
These values are very similar with the obtained value by our calculation. Thus, we presume that
the excitation of the $D$ meson be also describe by the potential with $\alpha=0.4$ and $k=0.9$ GeV/fm. 
In Fig.~\ref{fig:Bmeson} we show the calculated results for the $B$ and $B_{s}$ mesons
together with the observed spectra. We find that there is no strong discrepancy between 
the calculation and the observation. 


\subsection{Mass of diquark in $\Lambda_{c}$}

\begin{figure}
 \begin{center}
  \includegraphics[width=9cm]{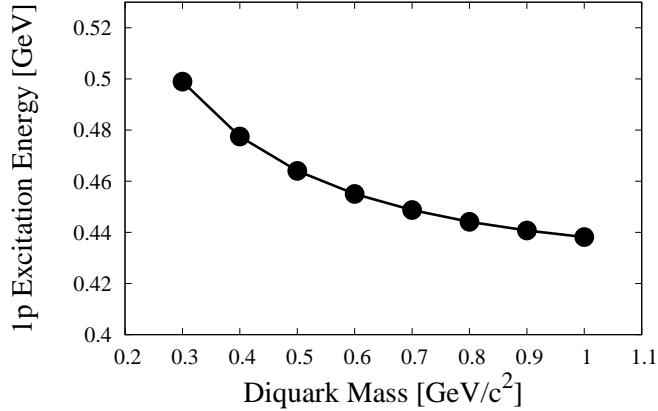}
 \end{center}
\caption{The $\Lambda_{c}$ excitation energy of the first excited state ($1p$) 
from the grand state ($1s$) as a function of the diquark mass with 
the $c$ quark mass $m_{c} = 1.5$ GeV/$c^{2}$ and 
the potential parameters $k=0.9$ GeV/fm and $\alpha=0.4$, which are
suitable for the meson systems.
The excitation energy for the observed $\Lambda_{c}$ is 0.330 GeV. 
The quark-diquark model with the confinement potential suitable for the meson systems and a reasonable 
diquark mass cannot reproduce the observed $\Lambda_{c}$ spectrum. }
\label{fig:md}
\end{figure}

With the success of the potential model for the meson spectra, we would make good use of the same
potential parameters for the $\Lambda_{c}$ spectrum.
Once we have fixed the potential parameters as $\alpha=0.4$ and $k=0.9$ GeV/fm
in the meson spectra, 
the diquark mass is only the parameter in the present model. 
The diquark mass can be fixed, for instance, by the 
excitation energy of $\Lambda_{c}$. In Fig.~\ref{fig:md}, we plot the diquark mass
dependence of the excitation energy of the first excited state ($1p$) from the grand state ($1s$).
The calculation is done without the spin-orbit interaction for the charm quark, 
which is responsible for the $LS$ splitting between $1/2^{-}$ and $3/2^{-}$.
In the range of the diquark mass from 0.3 GeV/$c^{2}$ to 1.0 GeV/$c^{2}$,
we obtain the excitation energy in the range of 0.5 GeV to 0.44 GeV. 
To compare these values with the observed $\Lambda_{c}$ excitation energies,
we take the spin weighted average of the excitation energies of the $1p$ states, $1/2^{-}$ and $3/2^{-}$.  
The matrix element of the spin-orbit interaction operator is calculated 
for the state with total angular momentum $J$, orbital angular momentum  $\ell$ and spin $s = 1/2$ as
\begin{equation}
  2 \langle J, \ell, 1/2 | \vec \ell \cdot \vec s |J, \ell, 1/2 \rangle =  J(J+1) - \ell(\ell+1) - \frac{3}{4}
  = \left\{ \begin{array}{rl} -2 \quad & {\rm for\ } J=1/2,\ \ell = 1 \\  1 \quad & {\rm for \ } J=3/2,\ \ell =1
    \end{array} \right. 
\end{equation}
Thus, if we take the weighted sum 
$E_{\rm ave} = \frac{2}{3} E_{3/2^{-}} + \frac{1}{3} E_{1/2^{-}}$,
we can remove the effect of the spin-orbit interaction in perturbation theory. 

The averaged value of the observed excitation energies of $\Lambda_{c}$ is found to be 0.330 GeV.
In the following, we refer to this value as experimental value of the $1p$ excitation energy of $\Lambda_{c}$.
The experimental value 0.330 GeV is out of the range of the calculated values 
with the diquark mass from 0.3 GeV/$c^{2}$ to 1.0 GeV/$c^{2}$, which is an expected range 
for the diquark as a bound state of two constituent quarks. 
Thus, we conclude that the confinement potential suitable for the meson systems cannot 
produce the $\Lambda_{c}$ spectrum in the quark-diquark picture with a reasonable diquark mass. 
This is a very interesting finding. The confinement force in the color electric interaction may be 
mainly dependent on the color charge and not on the flavor of the colored object. 
The present result, however, shows that the confinement force is different in the quark-antiquark and
quark-diquark systems in the case of the point-like diquark, 
although the antiquark and diquark have the same color charge. 
Therefore, as long as we take the quark-diquark picture, we should explain why the 
confinement potential is different in these systems.

\subsection{Potential parameters for $\Lambda_{c}$ and $\Lambda_{b}$} 

\begin{figure}
 \begin{center}
  \includegraphics[width=9cm]{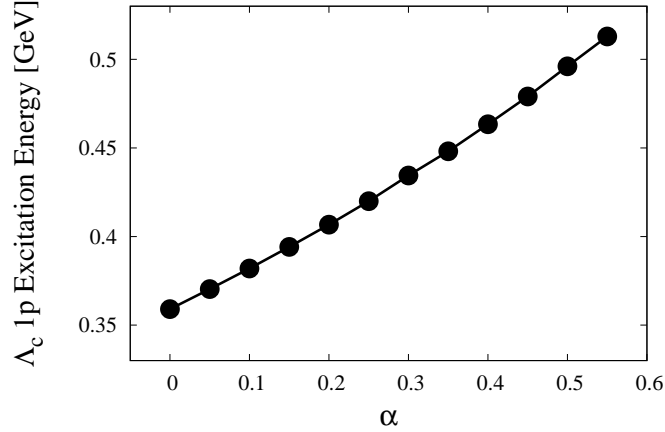}
 \end{center}
\caption{The $\Lambda_{c}$ excitation energy of the first excited state ($1p$) 
from the grand state ($1s$) as a function of $\alpha$
in the potential (\ref{pot}). 
The string tension is fixed so as to $k=0.9$ GeV/fm which is suitable for the meson spectra.
The masses of the $c$ quark and the diquark are 1.5 GeV/$c^{2}$ and 0.5 GeV/$c^{2}$, respectively. 
The excitation energy for the observed $\Lambda_{c}$ is found to be 0.330 GeV.  }
\label{fig:alpha}
\end{figure}

We have found that the confinement potential fitted by the quarkonium spectra does not 
reproduce the excitation energy of the $\Lambda_{c}$ baryon in the quark-diquark picture. 
In this section, we look for the potential parameters appropriate for the $\Lambda_{c}$ and
$\Lambda_{b}$ baryons. Let us fix the diquark mass as 0.5 GeV/$c^{2}$.
We have checked that we obtain qualitatively same results also for diquark mass 0.4 GeV/$c^{2}$.
 
\begin{figure}
 \begin{center}
  \includegraphics[width=9cm]{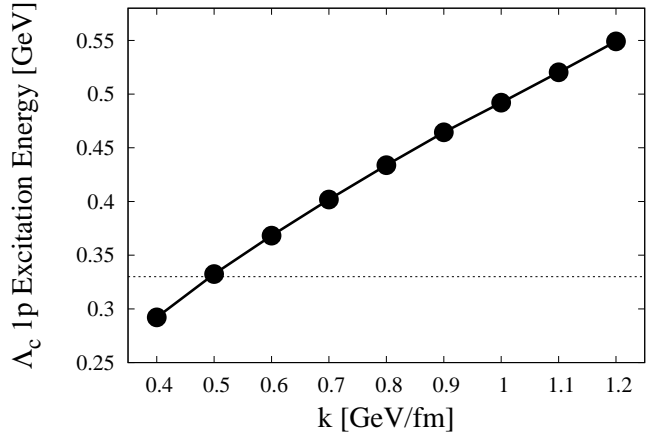}
 \end{center}
\caption{The $\Lambda_{c}$ excitation energy of the first excited state ($1p$) from the grand state ($1s$) 
as a function of the string tension $k$ in the potential (\ref{pot}).  The $\alpha$ parameter is fixed so 
as to $\alpha=0.4$.
The masses of the $c$ quark and the diquark are 1.5 GeV/$c^{2}$ and 0.5 GeV/$c^{2}$, respectively. 
The excitation energy for the observed $\Lambda_{c}$ is found to be 0.330 GeV, which is shown 
as the horizontal dotted line.  
The string tension around $k = 0.5$ GeV/fm reproduces the observed excitation energy. 
}
\label{fig:k}
\end{figure}

First of all, we see the $\alpha$ dependence of the $\Lambda_{c}$ excitation energy. 
In Fig.~\ref{fig:alpha}, we plot the excitation energy of the $1p$ state calculated
with various values of the $\alpha$ parameter. 
The string tension $k$ is fixed to be $k=0.9$ GeV/fm. 
The figure shows that the experimental excitation energy, 0.330 GeV, cannot be reproduced 
by $\alpha$ in the range of $\alpha =0.0$ to $0.6$ and $k=0.9$ GeV/fm.

\begin{figure}
 \begin{center}
  \includegraphics[width=10cm]{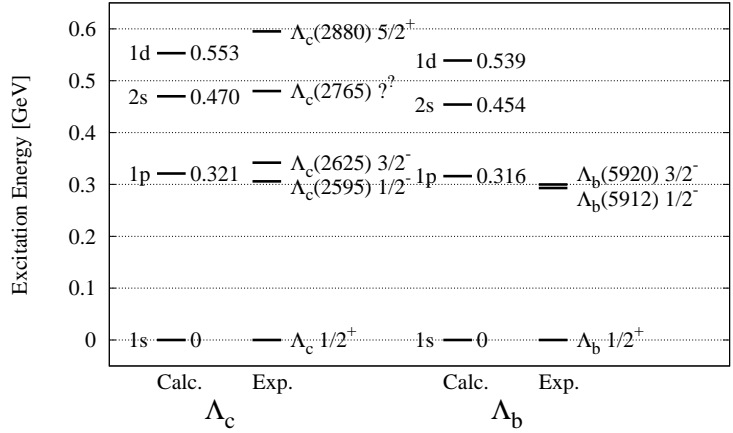}
 \end{center}
\caption{Comparison of calculated $\Lambda_{c}$ and $\Lambda_{b}$ excitation energies
from the grand state ($1s$) with the experimental data.
The potential parameters are fixed as $k=0.47$ GeV/fm and $\alpha=0.4$,
which reproduce the observed data well. 
The masses of the $c$ quark, the $b$ quark and the diquark are used as
1.5 GeV/$c^{2}$, 4.0 GeV/$c^{2}$ and 0.5 GeV/$c^{2}$, respectively. 
The experimental data are taken from Particle Data Group~\cite{PDG}. 
}
\label{fig:lambda}
\end{figure}

Next, we look for an appropriate value of the string tension $k$ for the $\Lambda_{c}$ spectrum. 
In Fig.~\ref{fig:k}, we plot the $1p$ excitation energies calculated with various values
of the string tension~$k$. The $\alpha$ parameter is fixed as $\alpha = 0.4$.
As we see in the figure, the experimental value, 0.330 GeV, can be reproduced 
with $k \sim 0.5$ GeV/fm. Calculating also the $\Lambda_{b}$ spectrum, 
we find that a best value of the string tension $k$ for both $\Lambda_{c}$ and $\Lambda_{b}$ 
is to be $k=0.47$ GeV/fm. Figure~\ref{fig:lambda} shows the comparison of 
the calculated results and observed spectra for $\Lambda_{c}$ and $\Lambda_{b}$,
suggesting that the potential with $\alpha=0.4$ and $k=0.47$ GeV/fm 
reproduces the whole excitation spectra of $\Lambda_{c}$ and $\Lambda_{b}$.
This value is substantially smaller than the string tension obtained for the meson systems,
$k=0.9$ GeV/fm. This implies that the confinement strength of the quark-diquark system
is almost half of that in the quark-antiquark system. This is again strange to universality for the confinement potential.

\begin{figure}
 \begin{center}
  \includegraphics[width=7cm]{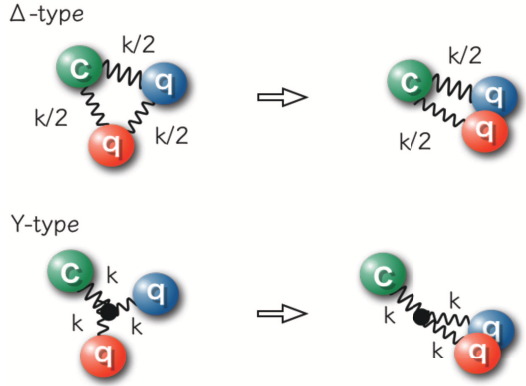}
 \end{center}
\caption{Schematic figure of interquark flux tube in a baryon and its diquark limit. There are two possibilities
of the gluon configuration in a baryon, $\Delta$-type and $Y$-type.  
}
\label{fig:diquarklimit}
\end{figure}

The weaker confinement potential for the quark-diquark system is unlikely to be 
explained by geometrical argument for interquark flux tube configuration, as long as 
we take a point-like diquark. 
As shown in Fig.~\ref{fig:diquarklimit}, there are two possibilities of the
flux configuration for three-quark baryons, 
so called $\Delta$-type~\cite{Cornwall:1996xr} and
$Y$-type~\cite{Capstick:1986bm,Takahashi:2000te}. 
In the $\Delta$-type configuration, three flux tubes between 
pairs of quarks form the triangle structure, while in the $Y$-type 
configuration there is one junction from which flux tubes are strung to the quarks.

According to the one-gluon exchange calculation,  
the attractive interaction between quarks with the $\bf \bar 3$ color configuration 
is half as strong as the interaction between a quark and an antiquark with
the color singlet configuration~\cite{Lucha:1991vn,Faessler:1982ik}. 
In the $\Delta$-type configuration, the string tension among the quarks 
may be given as $k/2$. In this case, if we take the diquark limit where 
two light quarks come to a same point, the string tension between the
heavy quark and diquark may be $k/2 + k/2 =k$ as shown in the upper part of 
Fig.~\ref{fig:diquarklimit}. This does not explain the reduction of the 
string tension in the quark-diquark system. In the $Y$-type configuration,
the string tension from the junction to the quark has a strength 
same as the quark-antiquark system with color singlet~\cite{Takahashi:2000te}.
If we take the diquark limit, the strength of the string tension between 
the heavy quark and diquark may be a value between $k$ to $2k$
depending on the position of the junction. For the junction 
at the diquark, the strength can be $k$, while the strength can be $k+k=2k$
if the junction located at the heavy quark. This does not again explain 
the weakness of the string tension for the quark-diquark system.

\subsection{Strange diquark mass}

\begin{figure}
 \begin{center}
  \includegraphics[width=9cm]{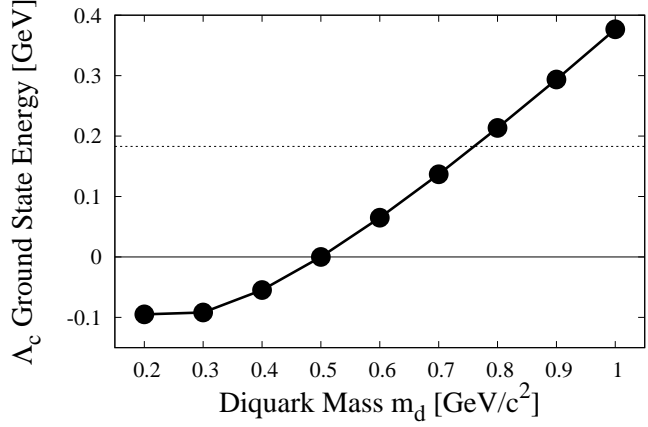}
 \end{center}
\caption{
Diquark mass dependence of the ground state energy of $\Lambda_{c}$ ($1s$)
measured from the $\Lambda_{c}$ mass with $m_{d} = 0.5$ GeV/$c^{2}$. 
The horizontal dotted line expresses the mass difference between 
the observed ground states of $\Lambda_{c}$ and $\Xi_{c}$. This figure suggests 
that the strange diquark mass $m_{ds} = 0.76$ GeV/$c^{2}$ reproduces
the observed $\Lambda_{c}$ and $\Xi_{c}$ mass difference. 
}
\label{fig:sdiquarkmass}
\end{figure}

Here we consider the scalar strange diquark which is composed of a light quark (up or down) 
and a strange quark forming the flavor and color antisymmetric configuration 
($\bf \bar 3$ both for the flavor and color spaces). 
The strange diquark can be a constituent of $\Xi_{c}$ ($\Xi_{b}$) together with
a charm (bottom) quark. 
Assuming the light flavor symmetry for the confinement potential~(\ref{pot}),
in which the potential is shared both for the $ud$ diquark and the strange diquark
up to the constant part of Eq.~(\ref{pot}), $V_{0}$, we can determine 
the strange diquark mass from, for instance, the mass difference between
$\Lambda_{c}$ and $\Xi_{c}$. 
In Fig.~\ref{fig:sdiquarkmass}, we show the diquark mass dependence
of the charmed baryon mass composed of a charm quark and a scalar diquark.
The charmed baryon mass is measured from the $\Lambda_{c}$ mass calculated with the diquark mass
$m_{d}=0.5$ GeV/$c^{2}$, where the $\Lambda_{c}$ spectrum is 
reproduced well. The experimental data for the difference 
between the $\Lambda_{c}$ mass and the isospin averaged
$\Xi_{c}$ mass is 0.18 GeV/$c^{2}$. This corresponds to the
calculation with the strange diquark mass $m_{ds}=0.76$ GeV/$c^{2}$,
which is a reasonable value for the strange diquark mass. 

With the strange diquark mass $m_{ds}=0.75$ GeV/$c^{2}$, 
we calculate the ground state $\Xi_{b}$ mass. 
We find that the mass difference between the ground states
of $\Lambda_{b}$ and $\Xi_{b}$ is calculated as 0.165 GeV/$c^{2}$.
We should compare this value with the experimental mass difference
0.175 GeV/$c^{2}$ and find that this is reasonably good agreement. 
For the ground state $\Xi_{b}$ mass, we take the isospin average.

\subsection{Excitation energy of $\Xi_{c}$}
\begin{figure}
 \begin{center}
  \includegraphics[width=10cm]{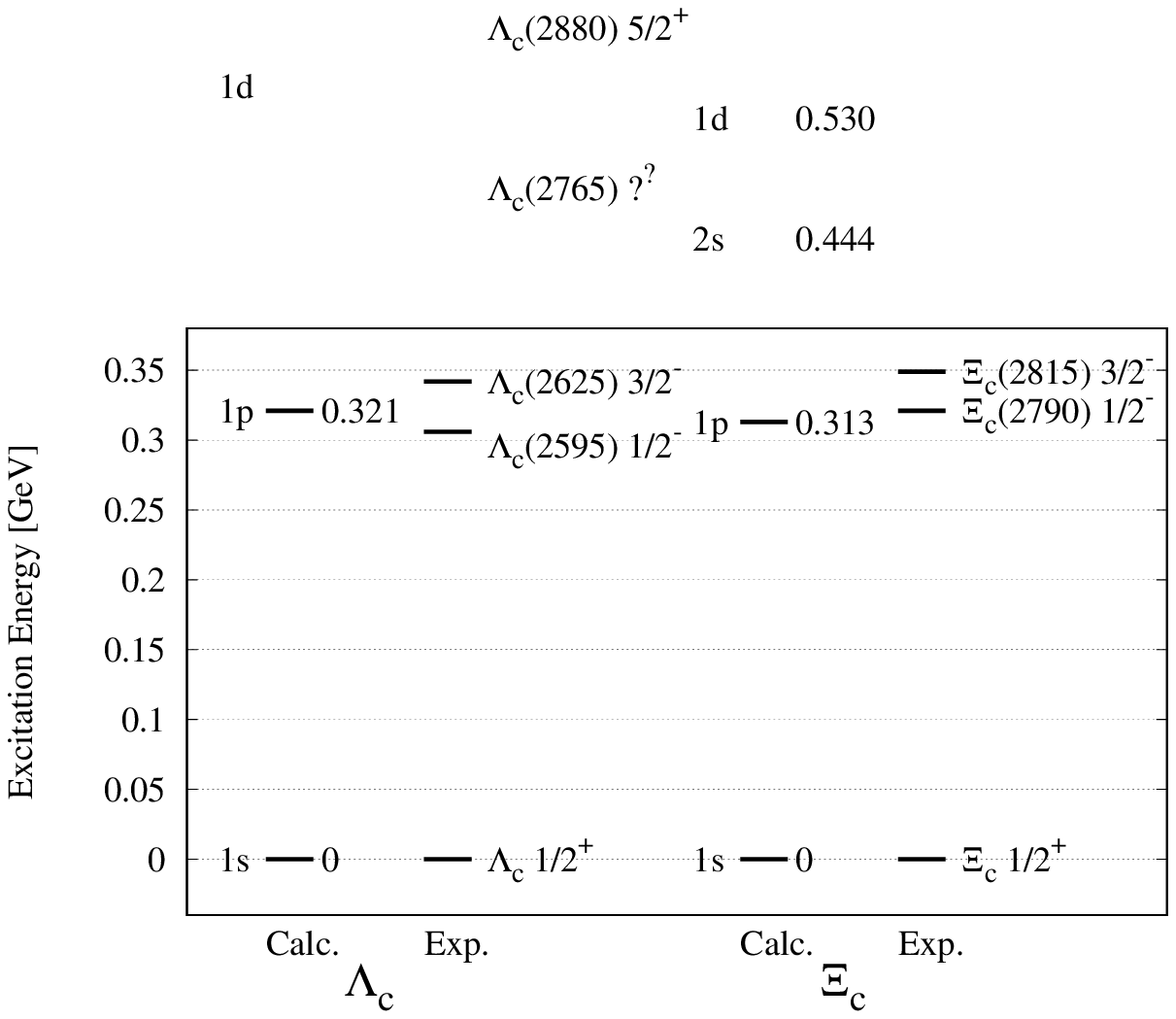}
 \end{center}
\caption{
Excitation energy of the $\Xi_{c}$ $1p$ state in comparison with $\Lambda_{c}$
and the observed excitation energies. 
The potential parameters are fixed as $k=0.47$ GeV/fm and $\alpha=0.4$,
which reproduce the observed $\Lambda_{c}$ data well. 
The masses of the $c$ quark and the strange diquark are used as
$m_{c}=1.5$ GeV/$c^{2}$ and $m_{ds}=0.76$ GeV/$c^{2}$, respectively. 
The experimental data are taken from Particle Data Group~\cite{PDG}. 
}
\label{fig:xic}
\end{figure}

\begin{figure}
 \begin{center}
  \includegraphics[width=9cm]{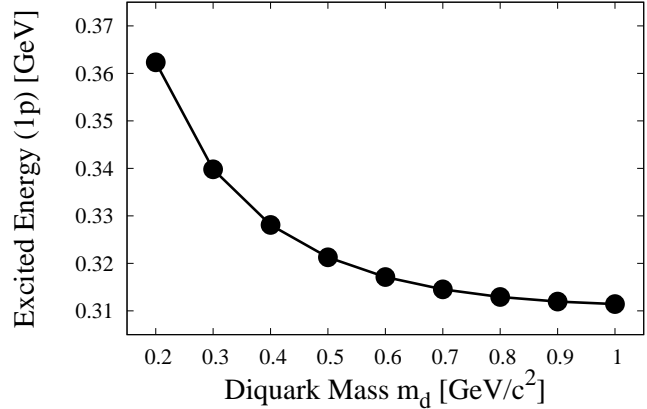}
 \end{center}
\caption{
Diquark mass dependence of the excitation energy of the $\Lambda_{c}$ $1p$ state
from the $1s$ ground state. 
This figure shows that we obtain the smaller excitation energy with the larger diquark mass.
}
\label{fig:excited}
\end{figure}

With the strange diquark mass $m_{ds}=0.76$ GeV/$c^{2}$,
we calculate the excitation energy of the $\Xi_{c}$ $1p$ state
from its $1s$ ground state. As shown in Fig.~\ref{fig:xic},
the excitation energy is obtained as 0.313 GeV.
Comparing the experimental data shown in the figure,
we find that the calculated value underestimates the 
observed masses of the $1p$ states. The size of the deviation 
from the observation is as small as the accuracy of the 
present model, and  we do not intend to discuss
this discrepancy further quantitatively. Nevertheless,
what puzzles us is that the calculated excitation energy 
of $\Xi_{c}$ is smaller than that of $\Lambda_{c}$,
while the experimental observation tells that they are reversed in order. 
As shown in Fig.~\ref{fig:excited}, in this potential model
the $1p$ $\Lambda_{c}$ state with the larger diquark mass has 
the smaller excitation energy, which fits our intuition that
the heavier object is harder to be excited. Unanticipatedly,
in nature the excitation energy of $\Xi_{c}$
is larger than that of $\Lambda_{c}$,
although $\Xi_{c}$ has a heavier component than $\Lambda_{c}$.
This is opposite to our model and intuition.
Thus, the present result is qualitatively against the observation. 
This disagreement could be solved by introducing the
mixing with the $\Xi_{c}^{'}$ state in which two light quarks
are symmetric in the flavor space. Nevertheless, the $\Xi_{c}^{\prime}$
state is a higher excited state than $\Xi_{c}$. Usually 
once the mixing of two states is introduced, level repulsion 
would take place and the lower state could be pushed down. 
This is an opposite direction to explain the smaller excitation 
energy of $\Xi_{c}$ than $\Lambda_{c}$.

\subsection{Decay width of $\Lambda_{c}$ excited state}
We play one more game with $\Lambda_{c}$ in the quark-diquark model with a point-like diquark.
We estimate the decay width of the $1p$ excited state of $\Lambda_{c}$. 
The decay widths of the observed $\Lambda_{c}$ states 
are as narrow as 2.6 MeV for $\Lambda(2595)$ with $1/2^{-}$ 
and less than 0.97 MeV for $\Lambda(2625)$ with $3/2^{-}$, 
which are extremely small in the strong decay. 
The main decay mode for both $\Lambda_{c}$ is 
$\pi\pi \Lambda_{c}$, and $\pi \Lambda_{c}$ is forbidden by isospin symmetry. 

In the flux tube model, 
pair creation of a quark and an antiquark induces  
the decay of a hadron into two lighter hadrons.
In the quark-diquark model for $\Lambda_{c}$, 
by the creation of a quark and an antiquark inside of the flux tube, 
the $\Lambda_{c}$ baryon decays into a $D$ meson and a nucleon, 
where the $c$ quark forms the $D$ meson 
with the created antiquark and the diquark forms the nucleon with the 
created quark. But, the $DN$ system has a larger rest mass than 
the $\Lambda_{c}$ excited state, so that it cannot decay into $DN$. 
With the diquark-antidiquark pair production, the $\Lambda_{c}$ excited state
decays into $\Lambda_{c}$ made of the charm quark and the created diquark
and a tetraquark formed by the created antidiquark and the diquark originally in the excited $\Lambda_{c}$,
and the tetraquark may be a sigma meson and decays into two pions. 

\begin{figure}
 \begin{center}
  \includegraphics[width=10cm]{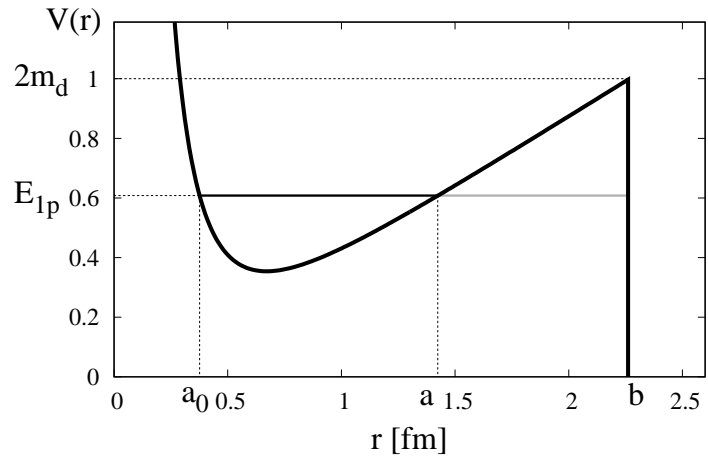}
 \end{center}
\caption{Potential~(\ref{tunnel}) with the centrifugal potential for $\ell=1$ as a function of $r$.
The energy $E_{1p}$ is the mass of $\Lambda_{c}$ in the $1p$ state 
measured from $m_{c}+m_{d}$.
The classical turning points are located at $a_{0}=0.38$ fm and $a=1.42$ fm.
The end of the tunnel is calculated as $b=2.27$ fm.
}
\label{fig:tunnel}
\end{figure}

Here let us estimate the decay width of $\Lambda_{c}$ in the $1p$ state
into $\Lambda_{c} \pi\pi$ by considering 
pair creation of diquark and antidiquark in the flux tube. 
We model this situation by taking the potential to be 
\begin{equation}
    V(r) = \left\{ \begin{array}{ll} - \frac{4}{3} \frac{\alpha}{r} \hbar c + kr + V_{0} & {\rm for}\ r<b \\
    0 & {\rm for}\ r>b \end{array} \right.  \label{tunnel}
\end{equation}
where $b$ is defined by the position where the potential energy reaches the energy for 
the diquark-antidiquark pair creation, $V(b) = 2m_{d}$. For larger distance than $b$, 
thanks to the pair creation of diquark and antidiquark, there are no interaction 
and the potential becomes zero.  The constant $V_{0}$ is determined 
so as to reproduce the absolute value of the $\Lambda_{c}$ ground state mass
by the potential~(\ref{pot})
together with the center of mass energy $m_{c} + m_{d}$. In the present case
$V_{0}=-0.039$ GeV. 
In Fig.~\ref{fig:tunnel}, we plot the potential~(\ref{tunnel}) together with 
the centrifugal potential for $\ell=1$. The energy $E_{1p}$ is the
mass of $\Lambda_{c}$ with $\ell=1$ measured from $m_{c}+m_{d}$.
The positions $a_{0}$ and $a$ are the classical turning points.
The excited $\Lambda_{c}$ state with $E_{1p}$ decays
by penetrating through the potential barrier from $r=a$ to $r=b$.
We calculate the decay width based on Gamov theory for nuclear $\alpha$ decay. 
There the tunneling probability is calculated by the WKB approximation:
\begin{equation}
  P = \exp\left[ - \frac{2}{\hbar c} \int_{a}^{b} \sqrt{2\mu c^{2} (V(r) - E_{1p})} dr \right]
  = 0.048.
\end{equation}
For the confined particle in the classical orbit,
the numbers of the trials of the penetration per unit time 
may be calculated as $v_{0}/(2R)$ with the length of the classical 
orbit $R=a-a_{0}$ and the velocity of the particle $v_{0}$.
We calculate the velocity by $v_{0} = \sqrt{\langle E_{\rm kin}\rangle /(2\mu)}$
with the expectation value of the kinetic energy $\langle E_{\rm kin}\rangle$
evaluated with the obtained wavefucntion,
$\langle E_{\rm kin}\rangle = 0.125$ GeV. 
Finally the decay width is estimated as
\begin{equation}
  \Gamma = \frac{\hbar}{\tau} 
  = \frac{\hbar c}{R} \sqrt\frac{\langle E_{\rm kin.} \rangle}{2\mu c^{2}} P
  = 3.8\ {\rm [MeV]},
\end{equation}
The size of the obtained value is consistent with the 
observed decay widths of the $1p$ states. It is noted that,
since the initial state with $\ell = 1$ has negative parity,
the relative angular moment between $\Lambda_{c}$
and the $\pi\pi$ system in the final state should be a odd number, such as $\ell =1$,
for $\pi\pi$ coming from a sigma meson. 
Nevertheless, 
in this model we do not consider spin and parity of the decay reaction.
Thus, we could have a further suppression factor for parity 
conserved configurations.


A similar calculation is also done using the potential with $k=0.9$ GeV/fm.
In the calculation we set the mass of the $1p$ state to be the observed mass,
which is not an eigenenergy of the potential with $k=0.9$ GeV/fm. 
We find the decay width
$\Gamma = 31$ MeV, which is one order of magnitude larger than the 
observed width. This fact also supports the smaller string tension 
for the heavy quark and diquark system. 
 
\subsection{Estimation of the diquark size effect}
As reported in Refs.~\cite{Alexandrou,Imai:2014yxa}, 
the size of the diquark can be as large as 1 fm.  Here we estimate the effect of the diquark size to
the excitation spectrum of $\Lambda_{c}$. 
Let us set the positions of the up, down and charm quarks to be ${\bm x}_{1}$, ${\bm x}_{2}$
and ${\bm x}_{3}$, respectively.
Assuming that the interaction strength of the inter quark force is half of the force between quark and antiquark,
we may write down the color electric potential between the charm quark and the light quarks as
\begin{equation}
   V_{\rm fs} = - \frac{4}{3} \hbar c \frac{\alpha}{2}
    \left( \frac{1}{|{\bm x}_{3} - {\bm x}_{1}|} + \frac{1}{|{\bm x}_{3} - {\bm x}_{2}|} \right) 
    + \frac{k}{2}   \left(  |{\bm x}_{3} - {\bm x}_{1}| + |{\bm x}_{3} - {\bm x}_{2}| \right) 
\end{equation}
where $\alpha$ and $k$ are the coupling strengths appearing in the quark-antiquark potential. 
Introducing a Jacobi coordinate defined by
${\bm \rho} = {\bm x}_{1} - {\bm x}_{2}$ and ${\bm r} =  {\bm x}_{3} - \frac{1}{2}({\bm x}_{1} + {\bm x}_{2}) $,
we write the potential as
\begin{equation}
   V_{\rm fs} = - \frac{4}{3} \hbar c \frac{\alpha}{2}
    \left( \frac{1}{|{\bm r} - \frac12{\bm \rho}|} + \frac{1}{|{\bm r} + \frac12{\bm \rho}|} \right) 
    + \frac{k}{2}   \left(  |{\bm r} - \frac12{\bm \rho}| + |{\bm r} + \frac12{\bm \rho}| \right) 
\end{equation}
We recover the potential (\ref{pot}) in the limit of ${\bm x}_{1}= {\bm x}_{2}$, that is ${\bm \rho} = 0$.
Assuming that two light quarks form a diquark, we do not regard $|{\bm \rho}|$
as a dynamical valuable rather as a parameter for the diquark size. 
The modulus of $\bm r$, $r = | {\bm r} |$, represents the distance between the diquark and the charm quark. 

Now we try two ways of the estimation of the diquark size effect. 
First we use perturbation theory by regarding the diquark size to be small. 
Let us expand the linear potential in terms of the diquark size $\rho$ and take the leading correction:
\begin{equation}
   \Delta V_{\rm fs}^{(1)} =  kr \frac{1-\cos^{2}\theta}{8} \frac{\rho^{2}}{r^{2}},
\end{equation}
where $\theta$ is the angle of ${\bm \rho}$ and $\bm r$.
We calculate the finite size effect for the $\Lambda_{c}$ excitation energy between the 1S and 1P states
as first order perturbation:
\begin{equation}
   \Delta E = \langle 1P | \Delta V_{\rm fs}^{(1)} | 1P \rangle - \langle 1S | \Delta V_{\rm fs}^{(1)} | 1S \rangle.
\end{equation}
To calculate the matrix elements, we use the wave functions obtained by solving the Sch\"odinger equation for
the unperturbed Hamiltonian with the original parameters for the quarkonia, $\alpha=0.4$ and 
$k=0.9$~GeV/fm, and we assume that $\bm \rho$ orients to the $z$ direction and take $\ell_{z} = 0$
for the 1P state. Then we find the finite size correction calculated by perturbation theory to be 
\begin{equation}
   \Delta E  = - \left(0.13\ {\rm GeV/fm}^{2} \right) \rho^{2}.
\end{equation}
This shows that the finite size effect reduces the excitation energy, 
and the magnitude of the correction is to be order of 0.1 GeV for the diquark size 1 fm, 
although 1 fm should not be a small enough value for perturbative calculation but
we could learn the order of magnitude of the size effect. In this calculation the angle 
between $\bm \rho$ and $\bm r$ is a dynamical valuable and 
we do not fix the orientation of the diquark against the charm quark direction. 

In the second estimation, we fix the angle between $\bm \rho$ and $\bm r$
to be orthogonal, but we solve the Schr\"odinger equation fully. With this assumption 
the potential is written as
\begin{equation}
   V_{\rm fs}^{(2)}(r) = - \frac{4}{3} \hbar c \frac{\alpha}{\sqrt{r^{2} + \frac{1}{4} \rho^{2}}}
    +  k \sqrt{r^{2} + \frac{1}{4} \rho^{2}}
\end{equation}
where $\rho$ is the parameter for the diquark size and this potential is a function of 
the distance between the charm quark and diquark, $r$. 
It can be understood from this potential that, for $r < \rho$, 
the charm quark is located between two light quarks and, consequently, 
color forces between the charm quark and the light quarks have opposite directions
and the forces are to be neutralized. Thus, around the center the color electric force 
between the charm quark and the diquark is less attractive that the case of the
point-like diquark. For $r > \rho$, $\rho$ would be negligible and we recover the original potential 
for the point-like diquark. 
Solving the Schr\"odinger 
equation with $\alpha = 0.4$ and $k=0.9$ GeV/fm,
we find the $\Lambda_{c}$ excitation energy to be 0.341 GeV for $\rho=0.5$ fm 
and 0.273 GeV for $\rho = 1.0$ fm, which should be compared with 0.464 
for $\rho = 0.0$ fm. Again we find that the finite size correction reduces 
the excitation energy and the diquark size 0.5 fm reproduces a reasonable
value for the excitation energy, which is consistent with the finding for
the finite size diquark in Refs.~\cite{Alexandrou,Imai:2014yxa}.
The reason of the suppression of the excitation energy for the finite size
diquark is the following; As discussed above, for $r < \rho$ the attraction
between the charm quark and the diquark gets weaker, and in such a case
the distribution of the $\Lambda_{c}$ wave function spread out
more than the case of the point-like diquark. This makes the centrifugal force 
less effective. Thus, the orbital excitation energy is reduced for the finite size diquark. 
For final conclusions, we should investigate the $\Lambda_{c}$
spectrum more precisely, for instance, by considering orbital and radial excited
states more and the $\Lambda_{b}$ spectrum. 
For quantitatively more precise argument, we should
perform three-body calculation by regarding three quarks as dynamical 
objects, which is, however, beyond the scope of this brief report. 
Here we just mention that the finite size effect can be a solution 
of the overestimation problem for the excitation energies, and 
that the diquark should have color polarization. 

\section{Summary and conclusion}
We have examined the excitation energy of the $\Lambda_{c}$ and
$\Lambda_{b}$ baryons in the quark-diquark model. In this model,
$\Lambda_{c}$ and $\Lambda_{b}$ composed of a heavy quark and 
a scalar $ud$ diquark with the $\bf \bar 3$ configuration for 
the flavor and color spaces, 
and these constituents are confined by the 
Coulomb and linear potential. 

We have found that 
the confinement potential suitable for the meson system
overestimates the $\Lambda_{c}$ and $\Lambda_{b}$
excitation energies, if we assume a point-like diquark.
We also find that, in order to reproduce the 
$\Lambda_{c}$ and $\Lambda_{b}$ spectra, 
the string tension should be half as strong as in the meson case.
This implies that if one takes the quark-diquark model for 
the $\Lambda_{c}$ and $\Lambda_{b}$ baryons with 
a point-like diquark, one should take such a smaller 
string tension for the quark-diquark system. 
This would give us an interesting question why the quark-diquark
system should have such a smaller string tension, 
because the color electric confinement 
potential is considered to be universal for the same color configuration.
We have estimated the decay width of the $\Lambda_{c}$ excited 
state based on Gamov theory of the nuclear alpha decay 
using the WKB approximation. We have found that the result obtained 
in this model is consistent with the experiment and this supports 
the smaller string tension for the quark-diquark model. 

The overestimation of the $\Lambda_{c}$ excitation energy
with the confinement potential suitable for the meson spectra
could be resolved by introducing the finite size for the diquark. 
In this case, for short distance of the heavy quark and the diquark,
the color forces between the heavy quark and the light quarks in the diquark
have opposite directions and are to be canceled. Thus 
the interaction between the heavy quark and the diquark for short distance 
is less attractive. This makes the relative distance of the quark and diquark larger 
in the $\Lambda_{c}$ wave function, and
the centrifugal force gets less effective.  
We have estimated the finite size effect to the 
$\Lambda_{c}$ excitation energy in two ways.
One is to use perturbation theory, and the other is to fix the
diquark orientation. Both calculations can be done within 
a simple two-body model. With these calculation, we have found that 
the finite size effect reduces the excitation energy and
the reduction magnitude would be 0.1 GeV to 0.2 GeV for the diquark size 1 fm.
This is a reasonable value to solve the overestimation. 
Thus, this finding can be one of the confirmations of the finite size diquark
in heavy baryons and implies that the diquark should have color polarization. 
For more precise argument for the diquark size, we certainly 
need full three body calculation.

We have also calculated the excitation energy of $\Xi_{c}$ as
a system of a charm quark and a strange diquark.
We have found that the excitation energy of $\Xi_{c}$ is 
moderately smaller that that of $\Lambda_{c}$. This is 
consistent with our intuition that a heavier particle is harder to be
excited, but it is inconsistent with the experimental observation 
in which the $\Xi_{c}$ excitation energy is slightly larger than
that of $\Lambda_{c}$. 

We hope that these findings give us a good chance to reconsider 
the nature of the confinement force in baryon for excitation spectra
of heavy baryons and the feature of diquark in heavy baryons. 

\section*{Acknowledgment}
One of the authors, D.J., thanks Prof.\ M.\ Oka and Dr.\ S.\ Yasui for useful discussion and suggestions. 
The work of D.J.\ was partly supported by Grants-in-Aid for Scientific Research from JSPS (25400254).
This work was done under Open Partnership with Spain of JSPS Bilateral Joint Research Project.


%


\end{document}